# Molecular beam epitaxy of superconducting $Sn_{1-x}In_xTe$ thin films

M. Masuko,[1, a)] R. Yoshimi,[2] A. Tsukazaki,[3] M. Kawamura,[2] K. S. Takahashi,[2] M. Kawasaki,[1, 2] and Y. Tokura[1, 2, 4]

[1)]*Department of Applied Physics and Quantum Phase Electronics Center (QPEC), University of Tokyo, Bunkyo-ku, Tokyo 113-8656, Japan.*

[2)]*RIKEN Center for Emergent Matter Science (CEMS), Wako, Saitama 351-0198, Japan.*

[3)]*Institute for Materials Research, Tohoku University, Sendai 980-8577, Japan.*

[4)]*Tokyo College, University of Tokyo, Bunkyo-ku, Tokyo 113-8656, Japan.*

We report a systematic study on the growth conditions of $Sn_{1-x}In_xTe$ thin films by molecular beam epitaxy for maximization of superconducting transition temperature $T_c$. Careful tuning of the flux ratios of Sn, In, and Te enables us to find an optimum condition for substituting rich In content ($x = 0.66$) into Sn site in a single phase of $Sn_{1-x}In_xTe$ beyond the bulk solubility limit at ambient pressure ($x = 0.5$). $T_c$ shows a dome-shaped dependence on In content $x$ with the highest $T_c = 4.20$ K at $x = 0.55$, being consistent to that reported for bulk crystals. The well-regulated $Sn_{1-x}In_xTe$ films can be a useful platform to study possible topological superconductivity by integrating them into the state-of-the-art junctions and/or proximity-coupled devices.

[a)]Electronic mail: masuko@cmr.t.u-tokyo.ac.jp

Novel physical properties arising from the non-trivial topology of band structures have been intensively studied in this decade.[1,2] Topological superconductivity (TSC) is one such topological state of matter which possesses a superconducting gap with a nonzero topological invariant in its interior and gapless excitations at the boundary.[3] The gapless excitations termed as Majorana fermions are theoretically predicted to obey non-Abelian statistics, and their applications to the quantum computation have been discussed.[3] To date, the presence of TSC has been reported in several systems,[4–17] which can be roughly classified into two groups. The first one is bulk superconductor materials intrinsically hosting topologically-nontrivial gaps.[4–9] The second one is the proximitized superconductor at the interface consisting of a conventional $s$-wave superconductor and a semiconductor with large spin-orbit interaction such as topological insulator and Rashba semiconductor.[10–17] By implementing the proximitized superconductor into device structures, possible signatures of Majorana fermions have been captured in the recent experiments.[10–15] This stimulates the demand for the intrinsic TSC in a thin film form compatible with the device fabrication, which would enable a wide variety of experiments controlling Majorana fermions.

SnTe is a topological crystalline insulator that possesses topological surface states protected by crystalline mirror symmetry.[18,19] Doped with In, $Sn_{1-x}In_xTe$ (SIT) exhibits superconductivity whose topological property has been vigorously debated mainly in bulk crystals.[8,9,20–31] While some spectroscopic studies point out the possible features of TSC such as zero-bias conductance peak and nontrivial surface states,[8,9,22,23] other experiments based on nuclear-magnetic-resonance, muon-spin spectroscopy, and heat capacity measurement indicate the conventional $s$-wave superconductivity.[21,28,29] In both cases, however, thin films of SIT would provide an intriguing platform for the observation of TSC. Even if SIT exhibits conventional superconductivity, the proximity effect may give rise to TSC in the heterostructures with telluride-based topological materials, for example topological insulator (Bi,Sb)$_2$Te$_3$ and topological crystalline insulator SnTe.[32,33] Moreover, although a unique dome-shaped $x$-dependence of superconducting transition temperature $T_c$ is demonstrated in bulk SIT crystals obtained by high-pressure synthesis,[25,26] such In-doping dependence has remained unexplored in thin films.[23,30] Systematic experiments on telluride-based superconducting thin films are of great importance to explore TSC in thin films or heterostructures.

In this study, we report a systematic study on the growth conditions and the electrical transport properties of SIT thin films. Fine-tuning of Sn, In, and Te flux ratios enables the growth of high-quality single-phase SIT thin films. We find that the In-content $x$ can be increased under the Te-reduced condition beyond the solubility limit of the bulk compound prepared under ambient pressure. $T_c$ shows dome-shaped $x$-dependence which is consistent with that reported for bulk SIT, indicating that the bulk comparable superconductivity of the SIT films can be implemented into heterostructures.

Thin films of SIT were grown on semi-insulating InP(111)A substrates using molecular beam epitaxy. Sn, In, and Te were supplied from Knudsen cells, and the equivalent beam pressures $P_{Sn}$, $P_{In}$, and $P_{Te}$ were monitored by a pressure gauge. Undoped SnTe was initially grown on the InP substrate at 400 °C for 5 minutes with $P_{Sn} = 5 \times 10^{-6}$ Pa and $P_{Te} = 1 \times 10^{-4}$ Pa, yielding an about 2 nm thick layer. The SIT film was subsequently grown at 460 °C for 60 minutes, corresponding to the deposition of approximately 25 nm evaluated by Laue fringes around the SIT (222)-diffraction. The total flux was fixed by maintaining $P_{Sn} + P_{In} = 5 \times 10^{-5}$ Pa for the fabrication of all the samples, whereas $P_{Te}$ was varied from $9 \times 10^{-6}$ Pa to $1.4 \times 10^{-4}$ Pa. Then we define $\beta = P_{Te} / (P_{Sn} + P_{In})$ as one of the critical control parameters for the SIT growth. In addition, we define $x$ based on the ratio of deposition time as $x = \frac{60}{65} \cdot \frac{P_{In}}{P_{Sn}+P_{In}}$, where $P_{Sn}$ and $P_{In}$ are the values under deposition of the SIT layer, by assuming following two points;

(1) deposition rates of SnTe and SIT are comparable, and (2) In is diffused to the bottom thin SnTe layer (~ 2 nm thick) so that the spatial distribution of In becomes homogeneous. The samples were characterized by the X-ray diffraction (XRD) with Cu $K\alpha 1$ radiation. The resistivity of the samples was measured by the standard four-terminal method. Electrical contacts were made by fixing the gold wires with silver paste on the sample surface. We used a $^3$He cooling system of a physical property measurement system (PPMS, Quantum Design) for the measurement down to 0.5 K under magnetic field. We also used an adiabatic demagnetization refrigerator system of PPMS for the measurement down to 0.1 K under zero magnetic field.

Figure 1(a) displays the XRD $2\theta$-$\omega$ scans for the SIT thin films with $x$ = 0, 0.28, 0.55, and 0.66, where $x$ = 0.66 is the largest In content obtained in this study. Clear Laue fringes appear at around both (111)- and (222)-diffraction peaks in the XRD patterns for all the samples, implying the high crystallinity of the films. The rocking curves around the diffraction peak of SIT (222) for all the films are reasonably sharp, whose full widths at half maxima are as narrow as 0.1º (Fig. 1(b)). In Fig. 1(c), the cubic lattice constant $a$ evaluated from the (222)-diffraction peak for all the films (red circles) is summarized as a function of $x$ for comparison with that for bulk values (gray circles).[25] The value of $a$ decreases monotonically with increasing $x$, which is consistent to that for the bulk samples. This consistent variation of $a$ supports the validity of our assumption regarding homogeneous In content in the whole films.

Figure 1(d) shows a parameter map of growth condition as a function of $x$ and $\beta = P_{\text{Te}} / (P_{\text{Sn}} + P_{\text{In}})$. In the low In-doping region below $x \sim 0.3$, the single-phase SIT films without secondary phase were obtained in a wide range of $\beta$ (red-shaded region with red circles and stars). Under $\beta \gg 1$, the excessive Te is expected to evaporate from the sample surface. As $x$ increases, however, $\beta$ needs to be finely tuned to suppress the segregation of secondary phases.

At the outer-region from the optimum red-shaded region, the segregation of secondary phase $In_2Te_3$ (blue squares and green triangles in Fig. 1(d)) and several impurity phases including tetragonal InTe (purple diamonds in Fig. 1(d)) were detected by XRD (Fig. 1(e)) for the Te-rich (larger $\beta$) and the Te-poor (smaller $\beta$) conditions, respectively. We found that only narrow conditions centered around $\beta = 4$ result in the single-phase SIT film with rich In content. The precise control of $\beta$ enables the growth of single-phase SIT films with rich In content up to $x = 0.66$ that exceeds the bulk solubility limit under ambient pressure ($x = 0.5$).[24] We speculate that SIT thin films with such rich In content are realized by a non-equilibrium crystallization process in molecular beam epitaxy, which sometimes enables the stabilization of metastable phases.

The temperature dependence of the resistivity for six films with different $x$ are presented in Fig. 2(a). Metallic temperature dependence of resistivity was observed in all films over a wide temperature range as shown in the inset. At low temperatures, the superconducting transitions appear as a sharp drop in resistivity for the films except for SnTe ($x = 0$). The $x$-dependence of transition temperature $T_c$ is plotted in Fig. 2(b). Here, $T_c$ is defined at the temperature where the resistivity reaches zero in the linear scale. The value of $T_c$ increases with $x$ and takes the highest value of 4.20 K at $x = 0.55$. This dome-shaped $x$-dependence of $T_c$ reproduces the reported trend in bulk samples,[24–27] which is mainly ascribed to the $x$-dependence of the density of states at the Fermi energy as clarified by theoretical calculation.[25,31] The similar $x$-dependence suggests the common origin for the superconductivity in bulk and thin films. Although the values of $T_c$ for the thin films are slightly lower than those for the bulk, we ascribe this difference to the film-thickness effect (see Sec. S1 in the Supplemental Material).

To evaluate the upper critical field $H_{c2}$ of the SIT film with $x = 0.55$, the magnetic field is applied in the out-of-plane and in-plane directions as depicted in Figs. 3(a) and (b), respectively (For the magneto-resistivity of the film with $x = 0.66$ and $x$-dependence of $H_{c2}^{\perp}$, see Sec. S2 in the Supplemental Material). Here, $H_{c2}$ is estimated from the onset temperature of superconducting transition under magnetic field, which is defined as the intersection between two lines that are extrapolated from normal ($\rho_{xx}^{N}$) and superconducting ($0.25$-$0.75\rho_{xx}^{N}$) regions. The discontinuity of data around 2.5 K in Figs. 3(a)-(c) is due to the different data sets measured at higher and lower temperatures than 1.8 K, the latter of which was taken after an interval of 10 days and subject to slight deterioration of the sample (for details, see Sec. S3 in the Supplemental Material). Although $H_{c2}$ in bulk crystals of SIT with cubic structure would be almost isotropic, $H_{c2}$ in a thin film is anisotropic under out-of-plane (Fig. 3(a)) and in-plane (Fig. 3(b)) magnetic field, reflecting the thin film form of the sample. As shown in the $H_{c2}$-$T$ plot in Fig. 3(c), while the temperature dependence of $\mu_0 H_{c2}^{\perp}$ is linear to $\left(1 - \frac{T}{T_c}\right)$, $\mu_0 H_{c2}^{\parallel}$ shows nonlinear $T$-dependence, which is typical of thin film superconductors.[34] In the framework of the two-dimensional (2D) Ginzburg-Landau (GL) model, the temperature dependence of $H_{c2}$ is described by $\mu_0 H_{c2}^{\perp}(T) = \frac{\phi_0}{2\pi(\xi(T))^2} \propto \left(1 - \frac{T}{T_c}\right)$ and $\mu_0 H_{c2}^{\parallel} = \frac{\sqrt{12}\phi_0}{2\pi\xi(T)t} \propto \sqrt{1 - \frac{T}{T_c}}$, where $\xi(T)$, $\phi_0$, and $t$ represent the GL in-plane coherence length, magnetic flux quantum ($h/2e$), and film thickness, respectively. In the case of out-of-plane $H$ in our experiment, we can estimate $\mu_0 H_{c2}^{\perp}(0) = 2.7$ T by extrapolating the linear fitting to 0 K, and thus $\xi(0)$ is evaluated to be $11.0 \pm 0.1$ nm, which is consistent with $\xi(0\text{ K}) = 15.0$ nm for bulk $Sn_{0.6}In_{0.4}Te$ in the literature.[27] On the other hand, $\mu_0 H_{c2}^{\parallel}(T)$ can be fit by $\propto \sqrt{1 - \frac{T}{T_c}}$. The value of $t$ giving the best fitting is $23.9 \pm 0.4$ nm, which is comparable to the thickness value evaluated from XRD (25 nm). It is noteworthy that, when the condition that $t \ll \xi$, which is assumed in the 2D GL model, is not satisfied, $H_{c2}^{\parallel}(T)$ is modified to be $\mu_0 H_{c2}^{\parallel} = \frac{\sqrt{12}\phi_0}{2\pi\xi(T)t}\left(1+\frac{9t^2}{\pi^6\xi^2}\right)$ due to the non-zero thickness effect.[34] In current situation, the factor $\frac{9t^2}{\pi^6\xi^2}$ is as small as 4 % at 0 K and even smaller at finite temperature, confirming the validity of our analysis based on the 2D GL model.

In summary, thin films of SIT were successfully grown in a wide range of In content $x$ by finely tuning the supplying ratio of Sn, In, and Te. By regulating the ratio of $P_{Te}$ / ($P_{Sn} + P_{In}$) at around 4, we successfully obtained fairly In rich SIT film ($x$ = 0.66) without segregation of secondary phases. The superconducting properties in our films almost reproduce those in bulk crystals; the dome-shaped $x$-dependence of $T_c$ and the in-plane coherence length. Besides, the thickness (25nm) of the present thin films is comparable with the in-plane coherence length to validate the analysis with the 2D GL model. Our result suggests that the thin films can be a useful platform to explore possible topological nature behind the superconductivity in SIT or SIT-based topological heterostructures.

We are grateful to M. Kriener and M. Uchida for useful discussions. This work was supported by JSPS/MEXT KAKENHI (Grant Numbers JP15H05853, JP17H04846, JP18H01155, JP18H04229, and JP19J22547) and JST CREST (Grant Numbers JPMJCR16F1 and JPMJCR1874).

**2**, 044802 (2018).

[26] K. Kobayashi, Y. Ai, H.O. Jeschke, and J. Akimitsu, Phys. Rev. B **97**, 104511 (2018).

[27] N. Haldolaarachchige, Q. Gibson, W. Xie, M.B. Nielsen, S. Kushwaha, and R.J. Cava, Phys. Rev. B **93**, 024520 (2016).

[28] S. Maeda, R. Hirose, K. Matano, M. Novak, Y. Ando, and G.Q. Zheng, Phys. Rev. B **96**, 104502 (2017).

[29] M. Saghir, J.A.T. Barker, G. Balakrishnan, A.D. Hillier, and M.R. Lees, Phys. Rev. B **90**, 064508 (2014).

[30] W. Si, C. Zhang, L. Wu, T. Ozaki, G. Gu, and Q. Li, Appl. Phys. Lett. **107**, 092601 (2015).

[31] T. Nomoto, M. Kawamura, T. Koretsune, R. Arita, T. Machida, T. Hanaguri, M. Kriener, Y. Taguchi, and Y. Tokura, Phys. Rev. B **101**, 014505 (2020).

[32] L. Fu and C.L. Kane, Phys. Rev. Lett. **100**, 096407 (2008).

[33] X.L. Qi, T.L. Hughes, and S.C. Zhang, Phys. Rev. B **82**, 184516 (2010).

[34] M. Tinkham, *Introduction to Superconductivity*, 2nd ed. (Dover, New York, 2004).

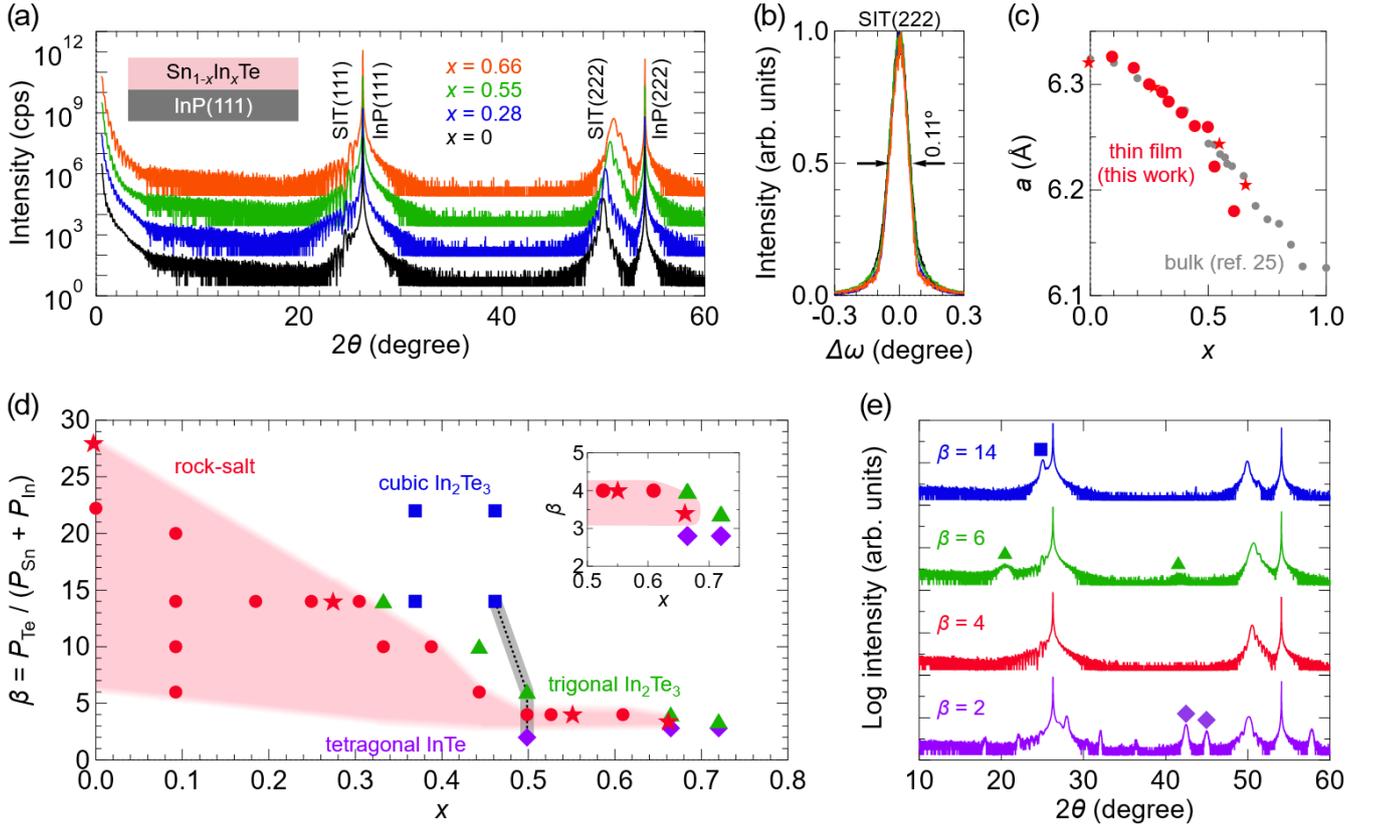

FIG. 1. (a) X-ray diffraction (XRD) patterns in $2\theta$-$\omega$ scans for $Sn_{1-x}In_xTe$ (SIT) thin films with various $x$. The inset schematically draws the sample structure. (b) Normalized rocking curves at (222)-diffractions for the samples shown in (a). (c) Cubic lattice constant calculated from the angle of (222)-diffractions (red circles) as a function of $x$. The data for bulk SIT taken from ref.[25] (gray circles) are also plotted for comparison. (d) A map of growth condition as a function of $x$ and the flux ratio $\beta = P_{Te} / (P_{Sn} + P_{In})$; $P_{Sn}$, $P_{In}$, and $P_{Te}$ being the equivalent beam pressure of the respective elements. The inset shows the magnified view around high-$x$ region. Red circles and stars, green triangles, blue squares and purple diamonds represent the rock-salt SIT, trigonal $In_2Te_3$, cubic $In_2Te_3$ and tetragonal InTe, respectively. In (c) and (d), red star symbols stand for the samples whose XRD patterns are exemplified in (a). (e) XRD patterns in $2\theta$-$\omega$ scans for the films with $x = 0.46 \sim 0.50$ grown under conditions with different $\beta$, which are highlighted in (d) with a gray line. Impurity peaks in XRD data are represented by the same symbols shown in (d).

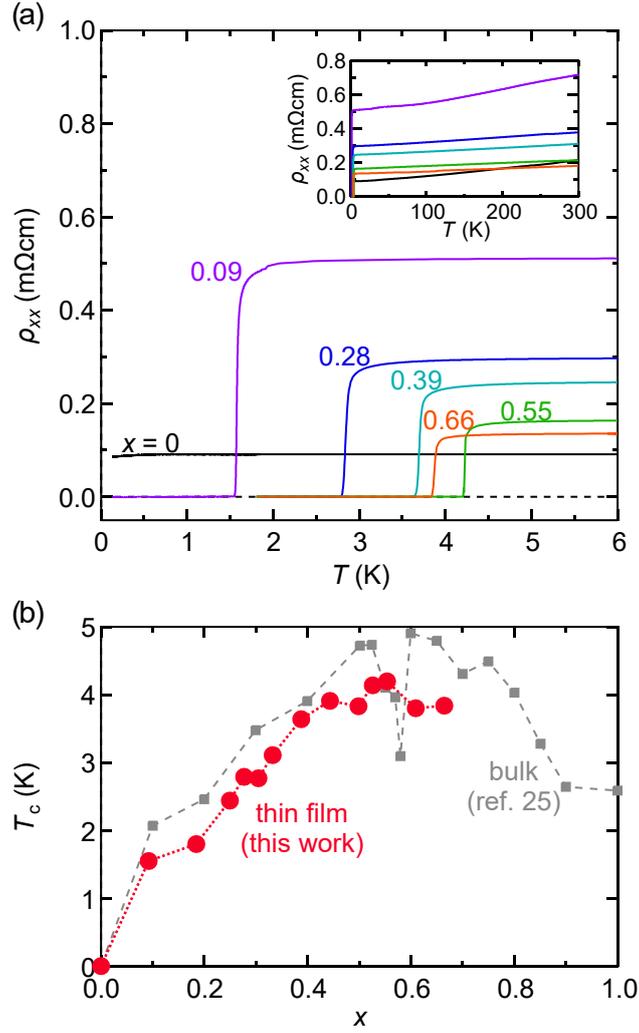

FIG. 2. (a) Temperature dependence of resistivity for SIT thin films with different $x$. The inset shows the resistivity up to 300 K. (b) $T_c$ as a function of $x$ for thin films (red) and bulk samples taken from ref.[25] (gray). In both cases, $T_c$ is defined at the temperature where the resistivity reaches zero in the linear scale.

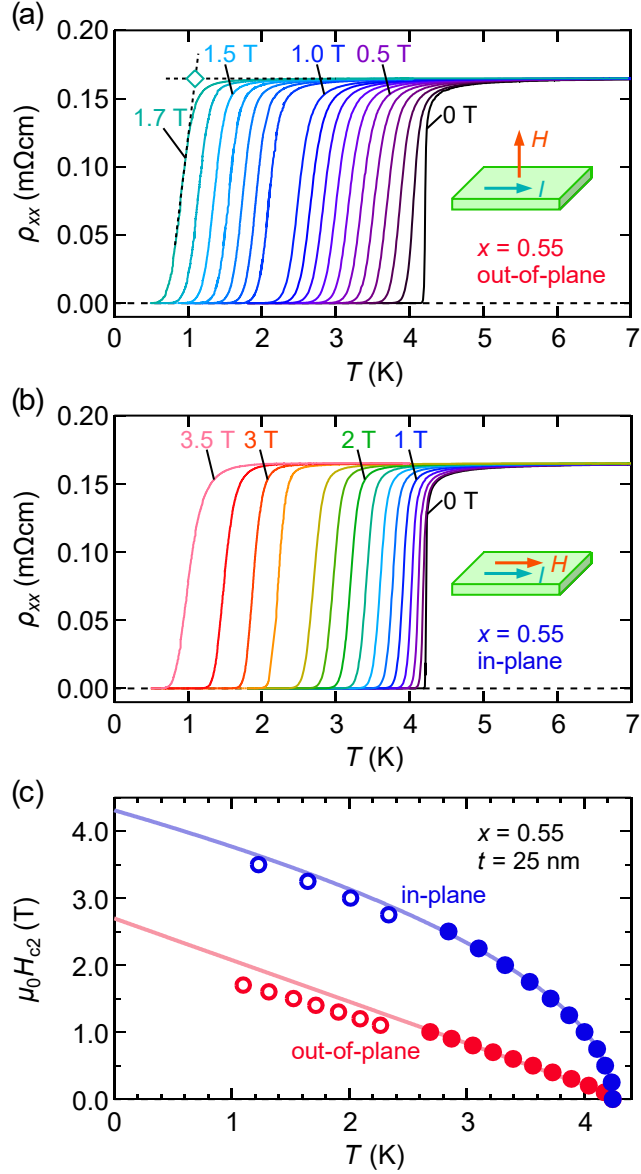

FIG. 3. (a) and (b) Temperature dependence of resistivity under (a) out-of-plane and (b) in-plane magnetic field for a Sn$_{0.45}$In$_{0.55}$Te ($x = 0.55$) thin film. The magnetic field is changed by 0.1 T for (a) and 0.25 T for (b). The insets schematically show the configuration of current $I$ and magnetic field $H$. An open diamond and dashed lines in (a) represent the definition of $\mu_0 H_{c2}$ in case of 1.7 T. (c) $\mu_0 H_{c2}$ as a function of temperature for out-of-plane (red) and in-plane (blue) magnetic field. The solid lines represent fitting curves (see the text). The closed and open circles are the data measured by the standard PPMS and a $^3$He cooling system, respectively. Only the former ones are used for the fitting (for details, see Sec. S3 in the Supplemental

Material).

Supplemental Material for

# Molecular beam epitaxy of superconducting $Sn_{1-x}In_x$Te thin films

M. Masuko,[1, b)] R. Yoshimi,[2] A. Tsukazaki,[3] M. Kawamura,[2] K. S. Takahashi,[2] M. Kawasaki,[1, 2] and Y. Tokura[1, 2, 4]

[1)]*Department of Applied Physics and Quantum Phase Electronics Center (QPEC), University of Tokyo, Bunkyo-ku, Tokyo 113-8656, Japan.*

[2)]*RIKEN Center for Emergent Matter Science (CEMS), Wako, Saitama 351-0198, Japan.*

[3)]*Institute for Materials Research, Tohoku University, Sendai 980-8577, Japan.*

[4)]*Tokyo College, University of Tokyo, Bunkyo-ku, Tokyo 113-8656, Japan.*

[b)]Electronic mail: masuko@cmr.t.u-tokyo.ac.jp

## Section S1 | Thickness dependence of $T_c$ and confinement effect

Figure S1 shows the temperature dependence of resistivity for Sn$_{1-x}$In$_x$Te (SIT) thin films ($x$ = 0.39) with various film thickness $t$. Here, the data for the 24-nm-thick film is identical to these in Fig. 2(a) in the main text. Thick films ($t$ = 41 and 84 nm) show superconducting transitions whose $T_c$ are comparable with that in a bulk single crystal (3.9 K). In contrast, thinner ones ($t$ = 7 and 12 nm), which have higher resistivity at normal states than thicker films, do not show zero resistivity above 1.8 K although the reduction in resistivity can be seen for the $t$ = 12 nm film. These data demonstrate that the reduction in thickness leads to the suppression of superconductivity. For the 24-nm-thick film, although $T_c$ is slightly suppressed compared with thicker ones, it shows zero resistivity at $T_c$ = 3.6 K.

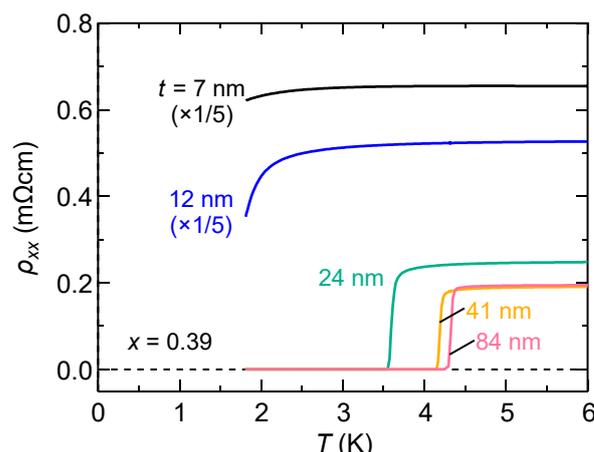

Fig. S1. Temperature dependence of the resistivity in SIT thin films ($x$ = 0.39) with various film thickness ($t$). For clarity, the resistivity of the 7- and 12-nm-thick films are divided by 5.

## Section S2 | Magneto-resistivity in the film with $x$ = 0.66 and $x$-dependence of $H_{c2}^{\perp}$

Figure S2(a) shows the magneto-resistivity of the film with $x$ = 0.66 at several temperatures. The transition from superconducting to normal state is induced by applying magnetic field, which takes place at a smaller magnetic field with elevating temperature. Figure

S2(b) displays the magneto-resistivity for several samples, in which the sharpness of the superconducting transition is not so sensitive to $x$ while $H_{c2}^{\perp}$ differs between samples. The $x$-dependence of $H_{c2}^{\perp}$ at 1.8 K is summarized in Fig. S2(c). Here, $H_{c2}^{\perp}$ is defined as the field at which the linear extrapolation of the superconducting transition reaches the resistance of normal state (see dotted lines in Fig. S2(b) that show the procedure to define $H_{c2}^{\perp}$ for $x = 0.28$). The dome-shaped $x$-dependence of $H_{c2}^{\perp}$ is observed, which is similar to that of $T_c$ (Fig. 2(b) in the main text). Owing to the thin-film growth of SIT with $x > 0.45$, it has been clearly demonstrated that $T_c$ as well as $H_{c2}^{\perp}$ shows dome-shaped $x$-dependence.

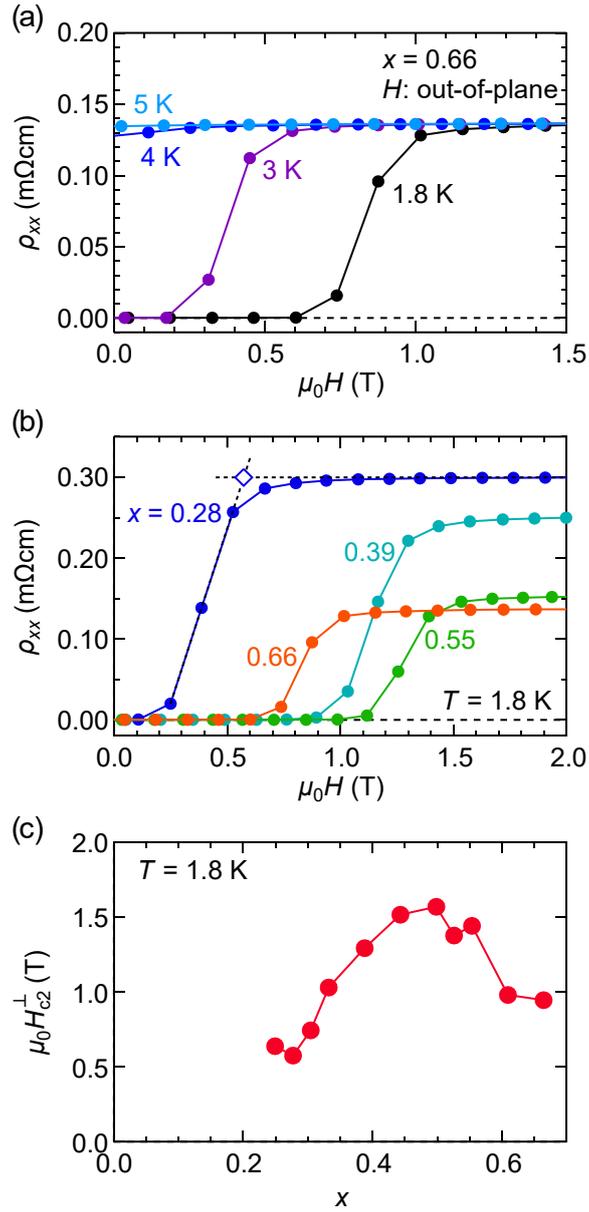

Fig. S2. (a) Magneto-resistivity under out-of-plane magnetic field in the thin film of Sn$_{0.34}$In$_{0.66}$Te ($x$ = 0.66) at various temperatures. (b) Magneto-resistivity under out-of-plane magnetic field at 1.8 K for the films with various $x$. An open diamond represents $\mu_0 H_{c2}^{\perp}$ for $x$ = 0.28 defined as the intersection between two broken lines extrapolated from normal and superconducting transition regions. (c) $\mu_0 H_{c2}^{\perp}$ at 1.8 K as a function of $x$, which is estimated from the $\rho_{xx}$-$H$ curves for each $x$.

**Section S3 | Decrease in $T_c$ over time**

The discontinuity in $\rho_{xx}$-$T$ curves between 1.0 T and 1.1 T (Fig. 3(a)) and between 2.5 T and 2.75 T (Fig. 3(b)) is lilely due to the deterioration of the sample occurring in the storage of the sample at room temperature for several days, which cannot be avoided by capping nor storing in a vacuum desiccator. Although the problem of deterioration remains, two sets of resistivity measurements were performed with the identical sample by different setup for the data set down to 0.5 K in addition to those above 1.8 K, resulting in the discontinuity of the data around 2.5 K. The measurement above 1.8 K was done by the standard temperature control system of the PPMS soon after the sample was fabricated. 10 days later, the second measurement down to 0.5 K was carried out by using a $^3$He cooling system. Because of the interval of the two measurements, $T_c$ of the film slightly decreased, causing the seeming discontinuity in $\rho_{xx}$-$T$ curves under magnetic field.

In Figs. S3 (a) and (b), we show the temperature dependence of $\rho_{xx}$ for bare and AlO$_x$-capped SIT ($x = 0.55$) films, respectively, to examine the degree of deterioration about $T_c$. The capping layer is 3-nm-thick AlO$_x$ fabricated by atomic layer deposition (ALD) at room temperature soon after removing the sample from the vacuum. We measured $\rho_{xx}$-$T$ curves three times, which were done on the day of the growth, 1 and 5 days after the growth. Both films were stored at room temperature in a vacuum desiccator between $\rho_{xx}$-$T$ measurements. $T_c$ is observed to decrease in both bare and capped samples by approximately 0.3 K after 5 days. We guess that the decrease in $T_c$ is not caused by the exposure to the air but the defect formation that occurs even at room temperature.

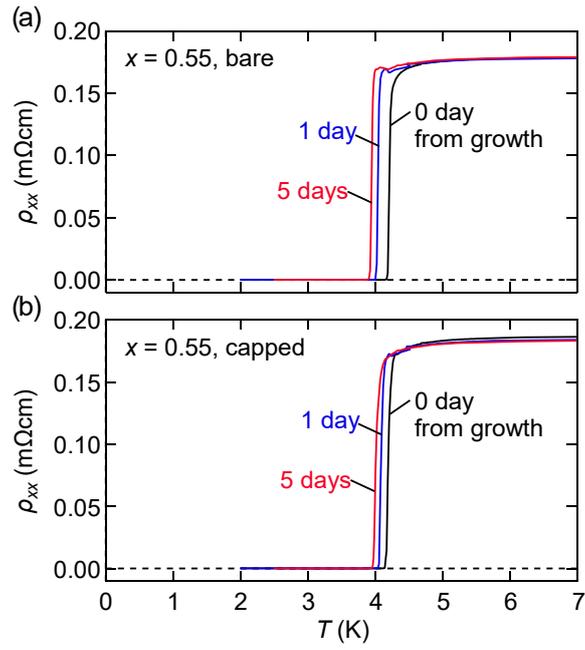

Fig. S3. (a) and (b) The temperature dependence of $\rho_{xx}$ for $x = 0.55$ in a bare film (a) and a film capped with $AlO_x$ (b), which were measured on the day of the growth, 1 and 5 days after the growth. The samples were stored in a vacuum desiccator at room temperature between measurements.